**Ion Flow under an Applied Electric Field in Semiconducting and Metallic Carbon Nanotubes**

J. B. Sokoloff, Northeastern University and Florida Atlantic University

Measurements made by Li, et. al., showed that the flow rates of potassium ions through *11nm* long, subnanometer diameter, metallic and semiconducting carbon nanotubes under an applied electric field are almost the same. In contrast, measurements of the electrical conductivity of potassium chloride solution through 100 $\mu m$ long metallic and semiconducting carbon nanotubes with diameters between 2.6-5.4nm by Cui, et. al., show that the ionic conductivity in the semiconducting nanotubes is larger than that of the metallic nanotubes. Possible theoretical origins of these differences are explored in this article.

## I. Introduction

Understanding the physics at the nanoscale for device applications, such as filtration devices [1-3], fuel cells[4-6] and sensing devices[7] requires an understanding of the flow of ions, protons and water molecules through confined narrow tube-like structures. An important aspect is to elucidate the role of electronic degree of freedom in determining the transport properties of carbon nanotubes (CNTs)[8].

In the short nanotubes in experiment reported in Ref. [9], the flow rate of potassium ions under an applied electric field is about the same in metallic and semiconducting nanotubes. This was explained in Ref. 10 by arguing that even the short semiconducting nanotube in these experiments at nonzero temperature has enough time to become an equipotential in an applied electric field, so that potassium ions in metallic and semiconducting nanotubes in an electric field have the same flow rate. In section II of this article, our argument is developed further. For the wider nanotubes in the experiments reported in Ref. 11, the ionic current consists of the flow of both potassium and chloride ions. Since the nanotubes discussed in Ref. 11 are 100 microns long, we show in section II that it will take a relatively long time for the semiconducting nanotubes to become equipotentials. In the experiments described in Ref. 11, the ionic conductivity is larger in semiconducting nanotubes. Since the metallic nanotube takes much less time to become an equipotential, there is no electric field inside the metallic nanotube, whereas there is an electric field inside the semiconducting nanotubes to push the ions through the nanotube. This is proposed as one possible explanation of the larger ionic conductivity in semiconducting nanotubes than in metallic nanotubes.

The fact that the flow rate in the experiments reported in Ref. 9 for protons and water molecules flowing through carbon nanotubes under osmotic pressure is greater in semiconducting nanotubes than in metallic nanotubes could, perhaps be because phonon friction is the dominant friction, and flow in these experiments is not caused by an applied electric field that can be cancelled out by conduction electrons in metallic nanotubes, as was proposed in Ref. 10.

## II. The difference between the electrical conductivity of a KCl solution in a metallic and a semiconducting nanotube

In this section, we will consider the concentration of conduction electrons in semiconducting and metallic carbon nanotubes, and use it to estimate how long it takes for a semiconducting and a metallic nanotube placed in an external electric field to become an equipotential. The concentration of electrons in a semiconducting nanotube is given by[12]

$$\begin{aligned} n_0 &= \int_0^\infty \frac{kdk}{2\pi}\left[\frac{1}{e^{\beta[\varepsilon(k)-\mu]}+1}+\frac{1}{e^{\beta[-\varepsilon(k)-\mu]}+1}\right] \\ &= \int_g^\infty \frac{\varepsilon d\varepsilon}{2\pi\hbar^2 v_F^2}\frac{1}{e^{\beta[\varepsilon-\mu]}+1}+\int_g^\infty \frac{\varepsilon' d\varepsilon'}{2\pi\hbar^2 v_F^2}\frac{1}{e^{\beta[\varepsilon'-\mu]}+1} \end{aligned} \quad (1)$$

where $\varepsilon = \varepsilon(k) = (\hbar^2 v_F^2 k^2 + g^2)^{1/2}$, $\varepsilon' = -\varepsilon$, and hence, for a semiconducting nanotube[12] with a gap *g*.

$$n_0 \approx \frac{1}{\pi\hbar^2 v_F^2}\int_g^\infty \varepsilon d\varepsilon e^{-\beta(\varepsilon-\mu)} = \frac{(k_B T)^2}{\pi\hbar^2 v_F^2}(\beta g+1)e^{-\beta(\varepsilon-\mu)} . \quad (2)$$

For a semiconducting nanotube with a band gap of 0.5eV, it is found from Eq. (2) that $n_0 = 3.44\times 10^7 m^{-2}$.

For a metallic nanotube, we use Eq. (1) for $n_0$ with $\beta g = 0$,

$$\begin{aligned} n_0 &= \frac{1}{2\pi(\hbar v_F)^2}\left[\int_0^\infty \varepsilon d\varepsilon \frac{1}{e^{\beta(\varepsilon-\mu)}+1}+\int_0^\infty \varepsilon' d\varepsilon' \frac{1}{e^{\beta(\varepsilon'-\mu)}+1}\right], \\ &= \frac{(k_B T)^2}{\pi(\hbar v_F)^2}\int_0^\infty \varepsilon'' d\varepsilon'' \frac{1}{e^{(\varepsilon''-\beta\mu)}+1} \end{aligned} \quad (3)$$

where $\varepsilon' = -\varepsilon$, $\varepsilon'' = \beta\varepsilon$, giving for the number of conduction electrons and holes per $m^2$, using the numerical result

$$I(\beta\mu = 0) = \int_0^\infty \frac{\varepsilon'' d\varepsilon''}{e^{\varepsilon''} + 1} = 0.82 \,,$$

$n_0 = 6.53 \times 10^{14} m^{-2}$. For $\mu \neq 0$,

$$n_0 = (6.53 \times 10^{14} m^{-2}) \frac{I(\beta\mu)}{0.82}$$

where *I* is the integral in Eq. (3) for $n_0$. The effective mass *m** for an undoped nanotube is given by

$$(m^*)^{-1} = \hbar^{-2} \frac{d^2\varepsilon(k)}{dk^2}\Big|_{k=0} = \left[ \frac{v_F^2}{(\hbar^2 v_F^2 k^2 + g^2)^{1/2}} - \frac{(\hbar^2 v_F^2 k)^2}{(\hbar^2 v_F^2 k^2 + g^2)^{3/2}} \right]_{k=0} = \frac{v_F^2}{g} \qquad (4)$$

giving $m^* = g / v_F^2 = 1.25 \times 10^{-31} kg$ for $g = 0.5 eV, v_F = 8 \times 10^5 m/s$.

As rough approximation, let us approximate the nanotube electric current density by the Drude model,

$$j = \frac{e\tau}{m^*} n_0 E \qquad (4)$$

where $\tau$ is the scattering time. The charge per unit length of a conducting nanotube is given by

$$\sigma(z) \approx \frac{4\pi\varepsilon_0 E z}{\ln 4\left(\frac{L^2 - z^2}{a^2}\right) - 2} \qquad (6)$$

which is found using the solution of the problem of the charge density of a metallic wire of radius a and length *2L* in an electric field *E*, as a function of the distance *z* along the wire in Ref. 13. It is plotted in Fig. 9. Since the charge of both a metallic wire and a metallic nanotube resides on their surfaces, Eq. (6) and figure 1 give the charge on the surface of the nanotube.

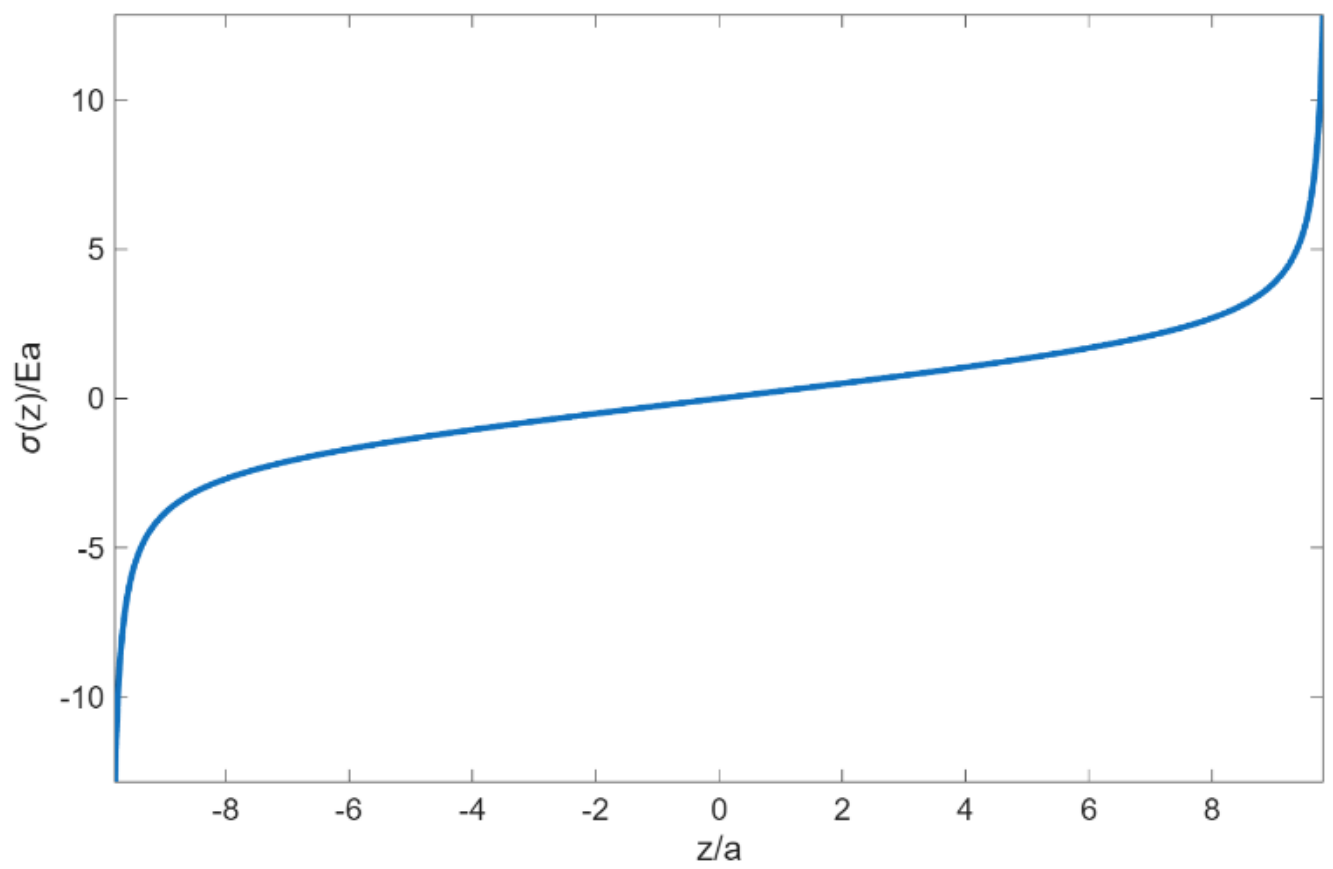

Figure 1: The electronic charge on the wall of a metallic nanotube $\sigma(z)/Ea$ in dimensionless units (assuming that *E* is in cgs units) is plotted as a function of *z/a*, the distance along the nanotube divided by the nanotube radius *a*, for *-9.8<z/a<9.8*.

The singularity when the denominator vanishes is integrable. Ref. 13 shows that the dipole moment of the wire in the field E is given by

$$P = \int_{-L}^{L} dz \frac{Ez^2}{\ln 4\left(\dfrac{L^2 - z^2}{a^2}\right) - 2} \approx \frac{EL^3}{3\ln\left(\dfrac{4L}{a}\right) - 7}. \quad (7)$$

Then, the charge that must flow along the nanotube to cancel out the field *E*, for *2L=11nm* and *a=0.4nm*, is given by

$$Q = \frac{P}{2L} = \frac{EL^2}{6\ln\left(\dfrac{4L}{a}\right) - 14} = 0.0735EL^2. \quad (8)$$

For the treatment of ion flow in nanotubes in Ref. 9, $2L = 10^{-8}m, a = 4\times10^{-10}m.$

Then, the time it takes to establish this charge is given by

$$-\int_0^t \frac{n_0 e^2 \tau}{m^*} E(t')C_h dt' \approx 4\pi\varepsilon_0 \frac{dQ}{dE} E(t), \quad (10a)$$

where $C_h$ is the circumference of the nanotube[12]. There is a minus sign because the electric field due to the polarization of the nanotube results in an electric field inside the tube in a direction opposite the dipole moment that results from the polarization of the nanotube. Taking the derivative with respect to *t*, we get

$$4\pi\varepsilon_0 \frac{dQ}{dE}\frac{dE(t)}{dt} = -\frac{n_0 e^2 \tau}{m^*} E(t) C_h \,, \tag{10b}$$

whose solution is

$$E(t) = E(0)e^{-t/t_0} \,, \tag{10c}$$

where

$$t_0 = \frac{4\pi\varepsilon_0 m^*}{n_0 e^2 \tau C_h}\frac{dQ}{dE} = 4.56\times 10^{-5} s \,, \tag{11}$$

for a semiconducting nanotube. For $t >> t_0$, *E(t)<<E(0)*.

If the mean free path $l_c = v_F \tau = 10nm, \tau = 1.25\times 10^{-14} s$. For a smaller value of $l_c, \tau$ will be smaller, and hence, $t_0$ will be larger. The correct value of $\tau$ can be determined by measuring the conductivity of the empty nanotube. Since for a metal, $n_0$ is larger by a factor of $10^{14}$, the time to establish the charge distribution on the tube $t_0$ is only $\approx 10^{-19} s$. So, both metallic and semiconducting nanotubes will become equipotentials rapidly. Since in the experiment reported in Ref. 9, each potassium ion is accompanied by a short chain of water molecules on each side, there can only be a single potassium ion within the width of the nanotube, as shown Ref. 9. Therefore, the ions travel through the nanotube as a line of ions. Since both the metallic and semiconducting nanotubes are equipotentials, the potassium ions are not acted on by the applied electric field once they enter the nanotube. Then, the motion of each ion that gets pushed into the nanotube by the electric field in the lipid droplet on one side of the nanotube is eventually stopped by the friction due to its interaction with the nanotube wall. The next ion to enter the tube transfers its momentum to the first ion, pushing it along in the tube. As this process repeats itself, there is a net flow of ions through the nanotube. This process is illustrated by a simple model in Appendix B of Ref. [10]. It is clear that the rate at which ions move through both the metallic and semiconducting nanotubes depends only on the rate at which ions enter the nanotube, independent of the strength of the friction due to the nanotube wall. Since both the semiconducting and metallic nanotubes are equipotentials, the flow rate of ions through the nanotube is independent of whether the tube is a metal or a semiconductor.

Since in Ref. 10, $2L = 10^{-4} m$, however, $t_0 = 1.00\times 10^3 s \approx 0.278 hr$ for the semiconducting nanotube. Thus, for $t << t_0$, the semiconducting nanotube in electric field will not be an equipotential region, which means that the ions inside the nanotube will be subject to the applied electric field inside the nanotube. Also, it will have fewer conduction electrons than the metallic nanotube, resulting in less friction due to the creation of electron excitations, and thus higher ionic conductivity. This implies that if one waits for a sufficiently long time, the semiconducting nanotube in Ref. 11 will become an equipotential, which suggests that the ionic conductivity of a metallic and semiconducting nanotube will become approximately equal, as they are in the experiment in Ref. 9. Since for a metallic nanotube $n_0$ is a factor of $10^7$ larger, $t_0$ will only be of the order of $10^{-3}s$, and therefore, it will become an equipotential very rapidly.

A sketch of the experimental set-up in Ref. 11 is given in Fig. 2.

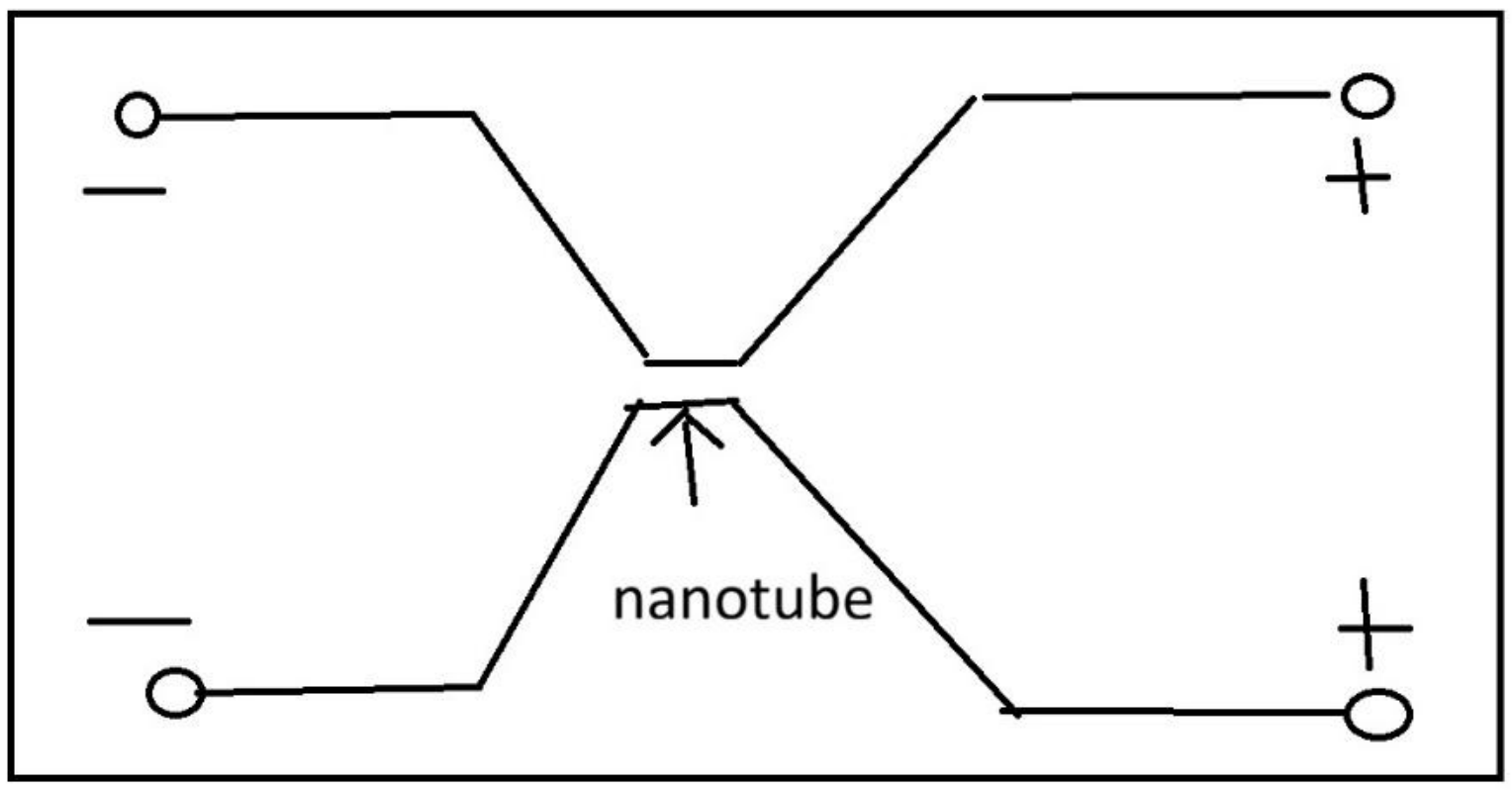

Fig. 2: A sketch of the experimental set-up in Ref. 11. Shown are the two positive and the two negative electrodes, which are poles perpendicular to the planar structure in which the fluid and the ions flow, and the etched guiding structures (illustrated in the sketch) that direct the ions into the nanotube.

Let us consider a simplified model of the set-up in Ref. 11, illustrated in Fig. 2, which is shown in Fig. 3, which illustrates the physics better.

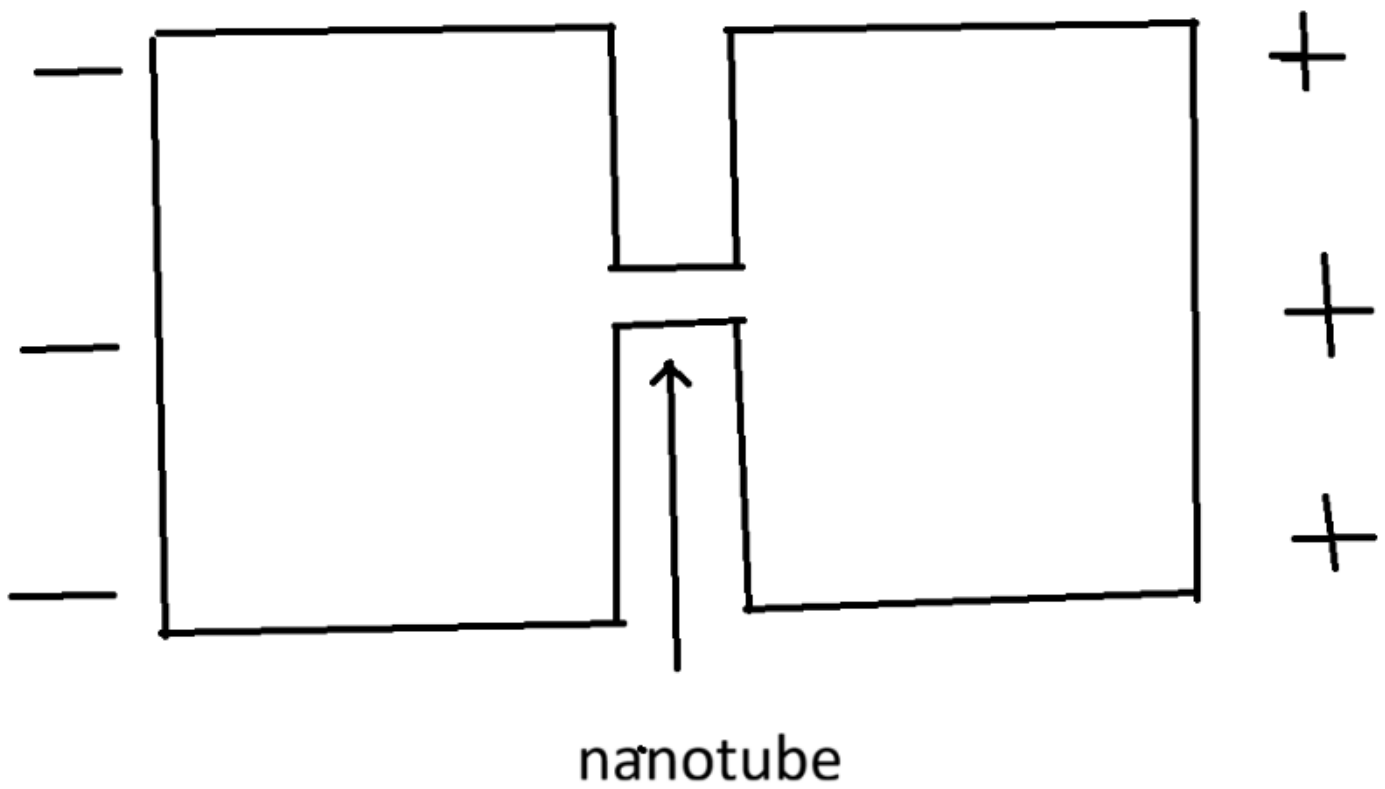

Fig. 3: Simplified model of the experiment reported in Ref. 11.

Let the potential difference between the two outer walls be *V* and the width of each box be *h*. In the experiments reported in Ref. 3, the length of the tube is negligible compared to *h*. Then, the electric field in the boxes is approximately *V/2h*. Then, the force exerted by the ions on the wall containing the mouth of the nanotube is equal to $Ahen_0(V/2h) = Aen_0V/2$ , where A is the area of the outer walls, and thus the pressure is $en_0V/2$. There is an equal negative pressure on the other side of the wall containing the nanotube, giving a net pressure $en_0V$ pushing the ions through the nanotube. In addition, inside a semiconducting nanotube for sufficiently short time, there is an electric field *V/2h*. Let $\lambda$ represent the friction coefficient for one of the ions, which includes friction from both the water and the wall. Then, for example, for the flow of a positive ion through a metallic nanotube,

$$\lambda v_1 = eV/h \qquad (12)$$

since the force pushing each ion through the tube is equal to eV/h, giving

$$J = n_0 e v_1 = n_0 e^2 V/(h\lambda) \ . \qquad (13)$$

(In reality, there is no force on the ions when they are inside the tube, but we assume that the forces on the ions when they enter the tube get transferred to the ions inside the tube via collisions.) In a semiconducting nanotube, there is an electric field inside the tube, which gives an additional contribution to the current,

$$n_0 e^2 V/(2h\lambda) \ , \qquad (14)$$

which can account for why the current density in the semiconducting nanotube is larger than that in the metallic nanotube.

## III. Conclusion

Whereas in Ref. 1 it was found that the rate of flow under an applied electric field of potassium ions in an 11nm long carbon nanotube with a 0.8 nm diameter is about the same in a semiconducting nanotube as in a metallic nanotube, it was reported in Ref. 3 that for 100micron long nanotubes with diameters between 2.6 and 5.4 nm, the flow rate of potassium chloride is faster in semiconducting than in metallic nanotubes. An explanation for this behavior was provided here for this difference based on the fact that whereas the shorter semiconducting nanotubes of Ref. 1 rapidly become equipotentials, the longer nanotubes in the experiments reported in Ref. 3 will require that the measurements be made over a time much longer than the *10s* over which the measurements reported in Ref.

11 were made[6] for the flow rates of the ions in semiconducting and metallic nanotubes to become nearly equal.

Acknowledgements: I wish to thank Ming Ma for many useful discussions of the results reported in Ref. 3.